\documentclass[journal=jpccck,manuscript=article]{achemso}
\usepackage[version=3]{mhchem} % Formula subscripts using \ce{}
\usepackage[T1]{fontenc}       % Use modern font encodings

\author{Weikang Wu}
\affiliation[SUTD]
{Research Laboratory for Quantum Materials, Singapore University of Technology and Design,  Singapore 487372, Singapore}

\author{Bo Tai}
\affiliation[SUTD]
{Research Laboratory for Quantum Materials, Singapore University of Technology and Design, Singapore 487372, Singapore}

\author{Shan Guan}
\affiliation[BIT]
{Beijing Key Laboratory of Nanophotonics and Ultrafine Optoelectronic Systems, School of Physics, Beijing Institute of Technology, Beijing 100081, China.}
\alsoaffiliation[SUTD]
{Research Laboratory for Quantum Materials, Singapore University of Technology and Design, Singapore 487372, Singapore}
\email{physguan@gmail.com}

\author{Shengyuan A. Yang}
\affiliation[SUTD]
{Research Laboratory for Quantum Materials, Singapore University of Technology and Design, Singapore 487372, Singapore}
\email{shengyuan_yang@sutd.edu.sg}

\author{Gang Zhang}
\affiliation[IHPC]
{Institute of High Performance Computing, Agency for Science, Technology and Research, 1 Fusionopolis Way, Singapore 138632, Singapore.}
\email{zhangg@ihpc.a-star.edu.sg}

\title
  {Hybrid Structures and Strain-Tunable Electronic Properties of Carbon Nanothreads}

\begin{document}

\begin{tocentry}

%\begin{figure}[!htb]
\centering
\includegraphics[width=0.75\textwidth]{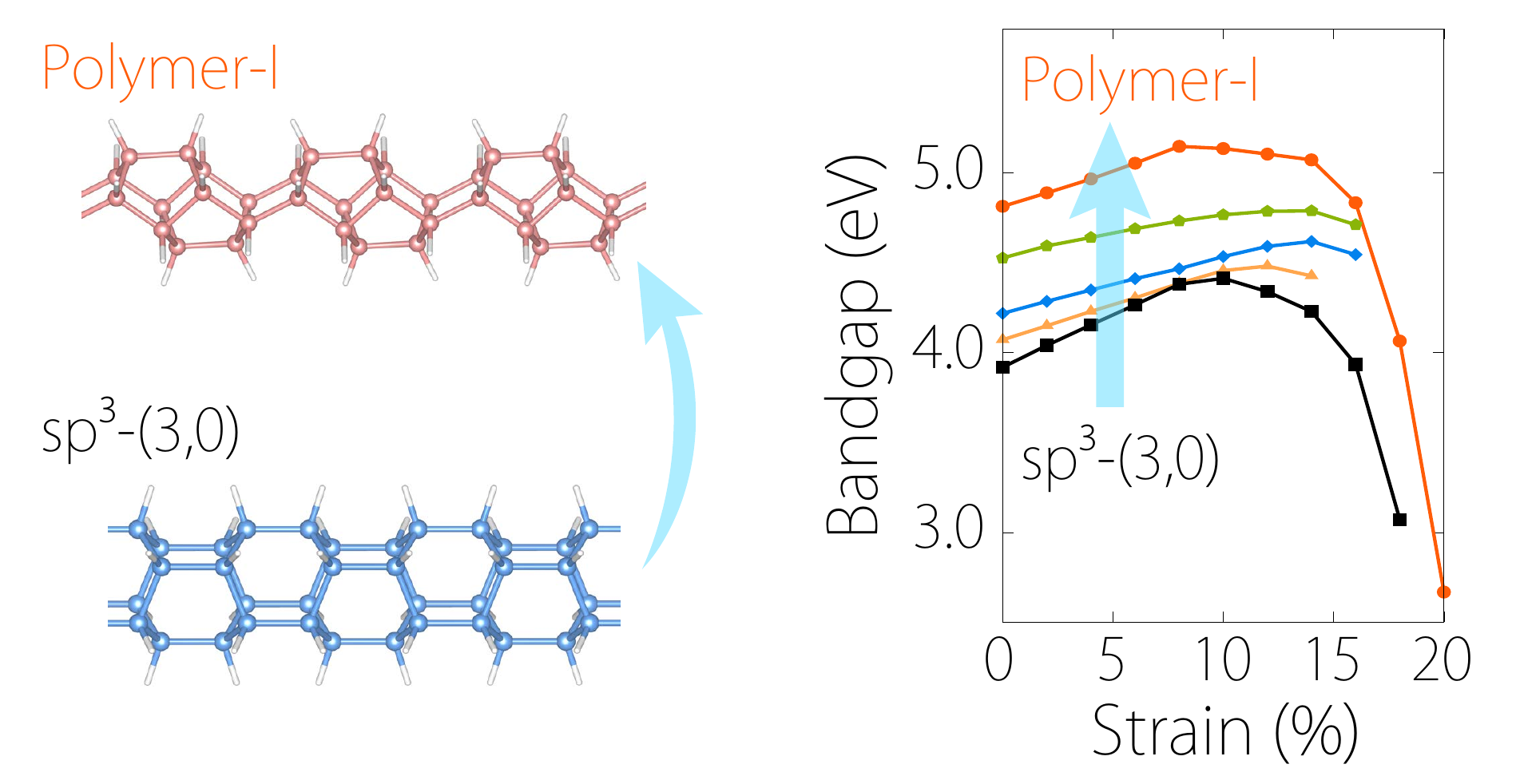}
%	%\caption{\label{TOC:bandgap}bandgap versus strain curves for the different types of DNTs}
%\end{figure}

\end{tocentry}

\begin{abstract}
The newly synthesized ultrathin carbon nanothreads have drawn great attention from the carbon community. Here, based on first-principles calculations, we investigate the electronic properties of carbon nanothreads under the influence of two important factors: the Stone-Wales (SW) type defect and the lattice strain. The SW defect is intrinsic to the polymer-I structure of the nanothreads and is a building block for the general hybrid structures. We find that the bandgap of the nanothreads can be tuned by the concentration of SW defects in a wide range of $3.92 \sim 4.82$ eV, interpolating between the bandgaps of $sp^{3}$-(3,0) structure and the polymer-I structure. Under strain, the bandgaps of all the structures, including the hybrid ones, show a nonmonotonic variation: the bandgap first increases with strain, then drops at large strain above 10\%. The gap size can be effectively tuned by strain in a wide range ($>0.5$ eV). Interestingly, for $sp^{3}$-(3,0) structure, a switch of band ordering occurs under strain at the valence band maximum, and for the polymer-I structure, an indirect-to-direct-bandgap transition occurs at about 8\% strain. The result also indicates that the presence of SW defects tends to stabilize the bandgap size against strain. Our findings suggest the great potential of structure- and strain-engineered carbon nanothreads in optoelectronic and photoelectrochemical applications as well as stress sensors.
\end{abstract}

\clearpage
\section{Introduction}
Carbon nanotubes have attracted tremendous interest, because of their extraordinarily small diameters, robust stability, and excellent electronic properties~\cite{tans1998room, de2013carbon, avouris2002molecular}. Remarkable observations have been made that the electronic properties of CNTs are sensitively determined by its chiral vector $(n, m)$ and the diameter $D$~\cite{kane1997size, ouyang2001energy, white1998density, hamada1992new, saito1992electronic, charlier2007electronic}, including metallic tubes ($n=m$), small-gap semiconductors ($n-m=3l$ with $l$ an integer) with gap size $\propto 1/D^{2}$, and insulating tubes with large gaps proportional to $1/D$. The electronic properties of carbon nanotubes may be strongly modified by the structural defects, which are often inevitable during fabrication processes~\cite{crespi1997situ, andriotis2002transfer, he2007effects, collins2000extreme}. Especially, one type of structural defects, the Stone-Wales (SW) defect\cite{stone1986theoretical}, has been found to exhibit many interesting effects. The SW defect consists of a pentagon-heptagon pair, which is induced by a $90^{\circ}$ rotation of a C-C bond in the hexagonal structures. Its presence can close the bandgap for large-gap nanotubes, open gaps for small-gap nanotubes, or increase the density of states for metallic ones~\cite{crespi1997situ, andriotis2002transfer}. Choi \emph{et al.} showed that SW defects can generate two quasi-bound states, leading to two quantized conductance reductions above and below the Fermi energy in metallic (10,10) nanotubes~\cite{choi1999ab, choi2000defects}. Furthermore, it has been demonstrated experimentally that the presence of SW defects can be readily controlled by mechanical distortion or by electron irradiation~\cite{nardelli1998mechanism,kotakoski2011stone}. Thus, the SW defects provide an alternative route to engineer the electronic properties of carbon structures.

Continuing efforts have been devoted to achieving tubes with smaller diameters. The motivation is partly from the expectation that the small nanotubes may exhibit interesting electronic properties due to the strong $\sigma^{*}$ and $\pi^{*}$ orbital mixing~\cite{blase1994hybridization}. The challenge is that for very small diameters (below 1 nm), the large curvature distortion typically causes the bond angles in the structure far below the ideal $120^{\circ}$ $sp^{2}$ angle, making the structure unstable. One possible solution is through attaching tightly-bonded groups (e.g., hydrogen or fluorine) to the structure, such that the $sp^{2}$-hybridized tube can be converted to a more stable $sp^{3}$-hybridized tube~\cite{Crespi2001tube30}.
Recently, such an ultrathin $sp^{3}$ carbon nanomaterial, called the carbon nanothread, has been successfully synthesized through the high-pressure solid-state reaction of benzene~\cite{Badding2015DNTs}. Most recently, carbon nanothreads with long-range order over hundreds of microns have been demonstrated through a mechanochemical synthesis method~\cite{li2017mechanochemical}. The word "nanothread" emphasizes its ultra-small diameter and the sp$^{3}$-bonding character of its carbon atoms is similar to that of diamond. It represents an ideal one-dimensional (1D) material, and its diameter is only about 6.4~\text{\AA}.
Theoretically, many topologically distinct structures are predicted for the nanothreads~\cite{Crespi2015DNTs}. Experimentally, from the derived pair distribution functions, the structure of carbon nanothreads were suggested to be a hybrid of two types of structures: the so-called $sp^{3}$-(3,0) structure~\cite{Crespi2001tube30} and the polymer-I structure~\cite{wen2011benzene}.
The $sp^{3}$-(3,0) nanothread can be viewed as a fully hydrogenated (3,0) carbon nanotube, while the polymer-I nanothread can be regarded as a 1D chain of SW defects with hydrogenated carbon atoms. In fact, starting from a pure $sp^{3}$-(3,0) nanothread, one can make a polymer-I nanothread by inserting the SW defects. The structures that interpolate between the two pristine types, i.e. nanothreads with $sp^{3}$-(3,0) domains connected by SW defects, may be termed as SW hybrid structures. Excellent mechanical and thermal properties have been proposed for these carbon nanothreads~\cite{Cranford2015DNTmecha, Gang2016DNTduct, Muniz2017DNTmecha, Gang2016DNTtensile, Gang2016DNTtherm}. For example, hybrid nanothreads are predicted to have a high stiffness of about 665-850 GPa and a large specific strength of $3.97 \times 10^{7}$ to $4.1 \times 10^{7}$ N$\cdot$m/kg, based on molecular dynamics (MD) simulations~\cite{Cranford2015DNTmecha}. And the structures show a brittle to ductile transition when the concentration of SW defects increases~\cite{Gang2016DNTduct}. Such properties make carbon nanothreads ideal for reinforced composites, strain sensors, and thermal connections~\cite{Gang2017DNTfibre, Gang2016DNTtherm}.

Besides mechanical and thermal properties, the electronic properties of carbon nanothreads also begin to attract attention. Preliminary studies have shown that carbon nanothreads are electrical insulators similar to diamonds~\cite{Crespi2001tube30}, and their electronic properties can be tuned through chemical functionalization~\cite{Muniz2017DNTfunctionalized}. In view of the important effects of SW defects as revealed in other carbon allotropes as well as the intrinsic presence of SW defects in polymer-I and hybrid nanothreads, it is of great interest to investigate how the SW defects modify the electronic properties of carbon nanothreads. Meanwhile, given the excellent mechanical properties of carbon nanothreads, it is natural to expect that strain may provide a powerful tool for tuning the electronic properties, as evidenced in the studies of other low-dimensional materials~\cite{minot2003tuning, levy2010strain, guinea2010energy,guan2015effects,wang2016strain}. In this work, we investigate these two issues using first-principles calculations. We focus on the pristine $sp^{3}$-(3,0) and polymer-I nanothreads and the SW hybrid nanothreads, which are the most relevant ones to experiment. 
We find that the bandgap of carbon nanothreads can be tuned by the SW defect concentration in a wide range as large as 1~eV, interpolating between the bandgaps of $sp^{3}$-(3,0) and polymer-I nanothreads. Under strain, the bandgaps of the nanothreads, including both pristine nanothreads and hybrid nanothreads, show a nonmonotonic variation. The bandgap widens with strain by about 0.5 eV up to a critical strain then dramatically drops with further increasing strain. Interestingly, for the $sp^{3}$-(3,0) nanothread, there is a switch of band ordering near the valence band maximum (VBM) with increasing strain; while for the polymer-I nanothread, there is an indirect-to-direct-bandgap transition occurring at about 8\% strain. Our work reveals that SW defects and strain offer powerful tools to control the electronic properties of carbon nanothreads, making them a promising platform for optoelectronic and photoelectrochemical applications.

\section{Computational Methods}
The first-principles calculations were performed based on the density functional theory (DFT) using the projector augmented wave method \cite{blochl1994PAW} as implemented in the Vienna \emph{ab initio} simulation package~\cite{kresse1993VASP, kresse1996VASP}. The generalized gradient approximation (GGA) with the Perdew-Burke-Ernzerhof (PBE) realization was adopted for the exchange-correlation potential~\cite{PBE1996PBE}. $1s^{1}$ and $2s^{2}2p^{2}$ were treated as the valence orbitals for H and C, respectively. The cutoff energy for the plane-wave basis was set to 500~eV, and the energy convergence criterion is set to be $10^{-5}$~eV. The cell parameters and the ionic positions are fully optimized until the residual force on each atom is less than $10^{-2}$~eV/\AA. For pristine $sp^{3}$-(3,0) and polymer-I nanothreads, a cubic cell of size $15~\text{\AA} \times 15~\text{\AA} \times c~\text{\AA}$ was used, which is large enough to eliminate the artificial interactions between the periodic images. To investigate the SW hybrid structures, $1 \times 1 \times 4$ supercells (with size $15~\text{\AA} \times 15~\text{\AA} \times 4c~\text{\AA}$) were employed to realize hybrid structures with different SW defect concentrations. The Brillouin zone was sampled with the $\Gamma$-centered $k$ mesh of sizes $1 \times 1 \times 11$ for the cubic cell, and $1 \times 1 \times 4$ for the supercell, respectively. Under strain, the atomic positions are fully relaxed without any constraint. To calculate the virial stress under strain, the cross-sectional area of the structure is approximated by $\lambda V_{0}$, where $\lambda$ is the linear atom density (shown in Table~\ref{tb1:data}) and $V_{0}$ is a reference atomic volume for carbon atom in bulk diamond, which is about $5.536$~\AA$^{3}$/atom. The same approach was previously used to characterize the (3,0) and (2,2) $sp^{3}$ tubes~\cite{Crespi2001tube30}. It should be noted that different approaches for calculating the cross-sectional area would yield different absolute values of the stress, but it will not affect the scaling behaviours as we focus on in this paper.

\section{Results and Discussion}

\begin{figure}[!htb]
	\includegraphics[width=0.8\textwidth]{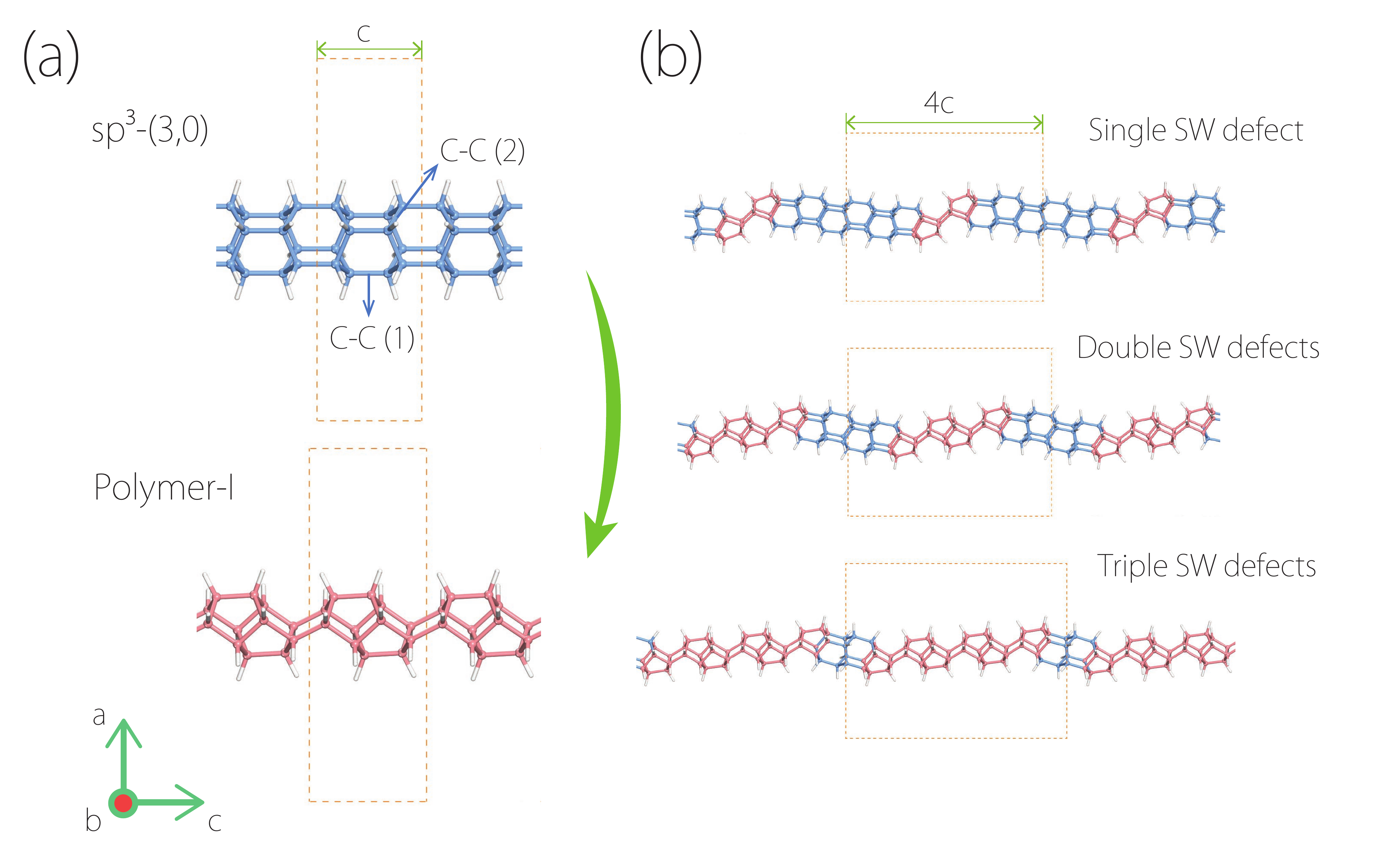}
	\caption{Structures of carbon nanothreads. (a) Structures of $sp^{3}$-(3,0) nanothread (upper panel) and polymer-I nanothread (lower panel). The C-C (1) and C-C (2) label the two types of C-C bonds in the $sp^{3}$-(3,0) nanothread. (b) SW hybrid nanothreads with single, double, and triple SW defects in a supercell. In the modelling, $1 \times 1 \times 4$ supercells are employed. Here the red color marks the SW defect regions.}
	\label{fg1:structures}
\end{figure}

The carbon nanothread models were constructed based on the recent experimental observations as well as first-principles calculations~\cite{Badding2015DNTs, Crespi2015DNTs}. The two pristine structures, the $sp^{3}$-(3,0) and the polymer-I structures, are shown in Fig.~\ref{fg1:structures}(a). One observes that the $sp^{3}$-(3,0) nanothread can be viewed as a fully hydrogenated (3,0) carbon nanotube without a SW defect, whereas the polymer-I nanothread can be regarded as a chain of SW defects. In the $sp^{3}$-(3,0) structure, there are two types of C-C bonds, C-C (1) and C-C (2)  [see Fig.1(a)], with the bond lengths of 1.569~\text{\AA} and 1.544~\text{\AA}, respectively. And the average C-C bond length for the polymer-I structure is about 1.550~\text{\AA}. These values are in good agreement with the experimental data ($\sim$1.52~\text{\AA})~\cite{Badding2015DNTs}. By generating SW defects in the $sp^{3}$-(3,0) structure, one can make the SW hybrid structures [see Fig.~\ref{fg1:structures}(b)] that interpolate between the two pristine structures. In this work, we adopt a $1 \times 1 \times 4$ supercell, so the number of SW defects in a single supercell ranges from 1 to 3, as shown in Fig.~\ref{fg1:structures}(b). Each SW hybrid nanothread consists of two distinct sections: the $sp^{3}$-(3,0) domain and the SW defect domain. 

\begin{table}
	\caption{The linear carbon atom density $\lambda$, energy per (CH)$_{6}$ unit, and bandgap for the SW hybrid nanothreads. The energies are relative to a single sheet of graphane.}
	\label{tb1:data}
	\begin{tabular}{llll}
		\hline
		& $\lambda=N/c$ (C atom/\text{\AA})  &  Energy (eV/(CH)$_{6}$)  &  Bandgap (eV) \\
		\hline
		$sp^{3}$-(3,0)         & 2.792  &  0.710  &  3.92  \\
		single SW   & 2.733  &  0.719  &  4.07  \\
		double SW  & 2.637  &  0.737  &  4.22  \\
		triple SW  & 2.525  &  0.759  &  4.53  \\
		Polymer-I                   & 2.410  &  0.795  &  4.82  \\
		\hline
	\end{tabular}
\end{table}

From our calculation, we find that the SW defects tend to stretch the length of the nanothreads, which can also be observed by the decreasing linear carbon atom density ($\lambda$) with the increasing number of SW defects (see Table~\ref{tb1:data}). For the $sp^{3}$-(3,0) nanothread, $\lambda = 2.792$~atoms/\text{\AA}, in accordance with the previous theoretical work~\cite{Crespi2015DNTs}. And the polymer-I nanothread exhibits the lowest linear density. As the concentration of SW defects increases from $0\%$ to $100\%$, the energy per (CH)$_{6}$ unit exhibits a continuing increase, similar to the previous MD simulation result which shows that adding a SW defect would increase the potential energy of carbon nanothreads~\cite{Cranford2015DNTmecha}.

\begin{figure}[!htb]
	\includegraphics[width=0.8\textwidth]{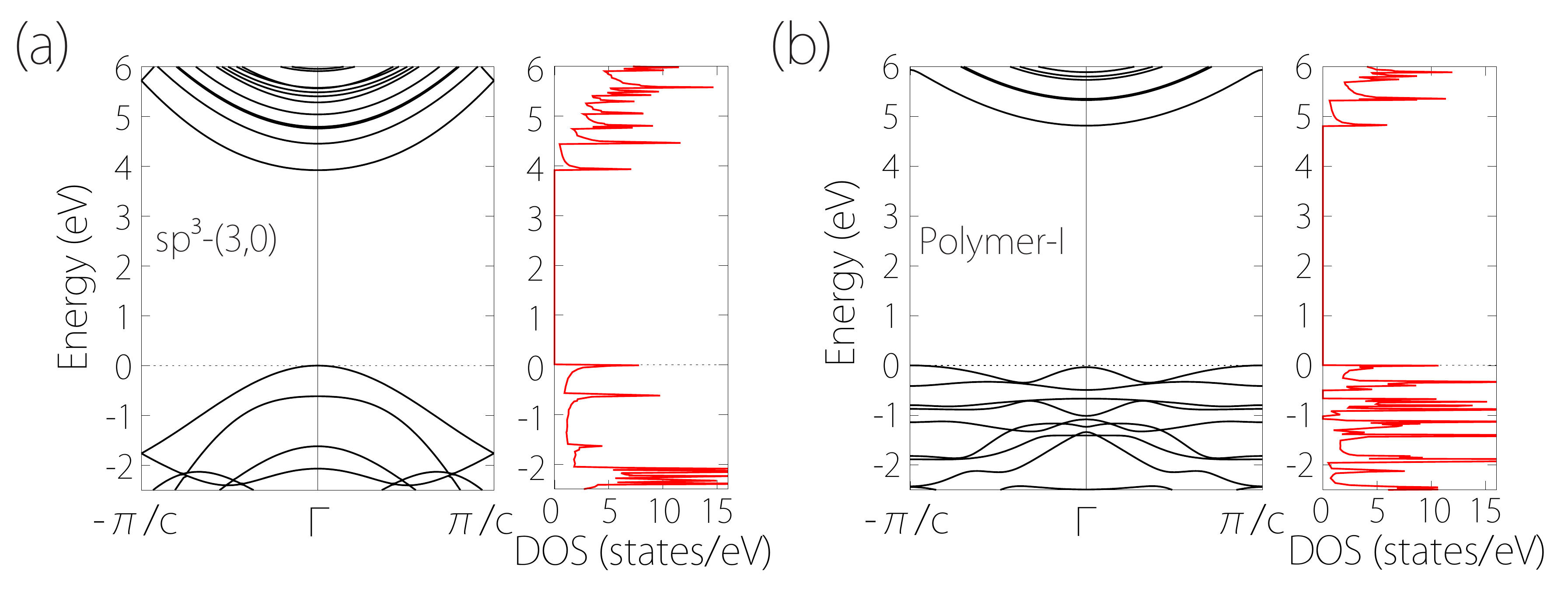}
	\caption{Band structures along with density of states (DOS) for (a) $sp^{3}$-(3,0) nanothread and (b) polymer-I nanothread.  The Fermi level is set at valence band maximum (VBM).}
	\label{fg2:band1}
\end{figure}

The calculated band structures of $sp^{3}$-(3,0) and polymer-I nanothreads along with the density of states (DOS) are shown in Fig.~\ref{fg2:band1}. Due to the $sp^{3}$ C-C bonding with saturated hydrocarbons, both nanothreads are wide-bandgap semiconductors, with gap value exceeding those obtained in $sp^{2}$ carbon nanotubes~\cite{hamada1992new}. The $sp^{3}$-(3,0) nanothread has a direct bandgap of about $3.92~eV$ at the $\Gamma$ point, whereas the polymer-I nanothread has an indirect bandgap with a larger value of $4.82$~eV, close to the bandgap values in the previous theoretical work~\cite{Crespi2015DNTs}. For the polymer-I nanothread, the conduction band minimum (CBM) is at the $\Gamma$ point, but the VBM is located at the Brillouin zone boundary. Comparing with the VBM in $sp^{3}$-(3,0) nanothreads, the much flatter bands near the VBM in the polymer-I nanothread show much larger hole effective mass and a more localized character. This feature is also reflected in the DOS for the two structures: One clearly observes that the DOS for polymer-I has a more pronounced peak near VBM in comparison with that for $sp^{3}$-(3,0).

\begin{figure}[!htb]
	\includegraphics[width=0.8\textwidth]{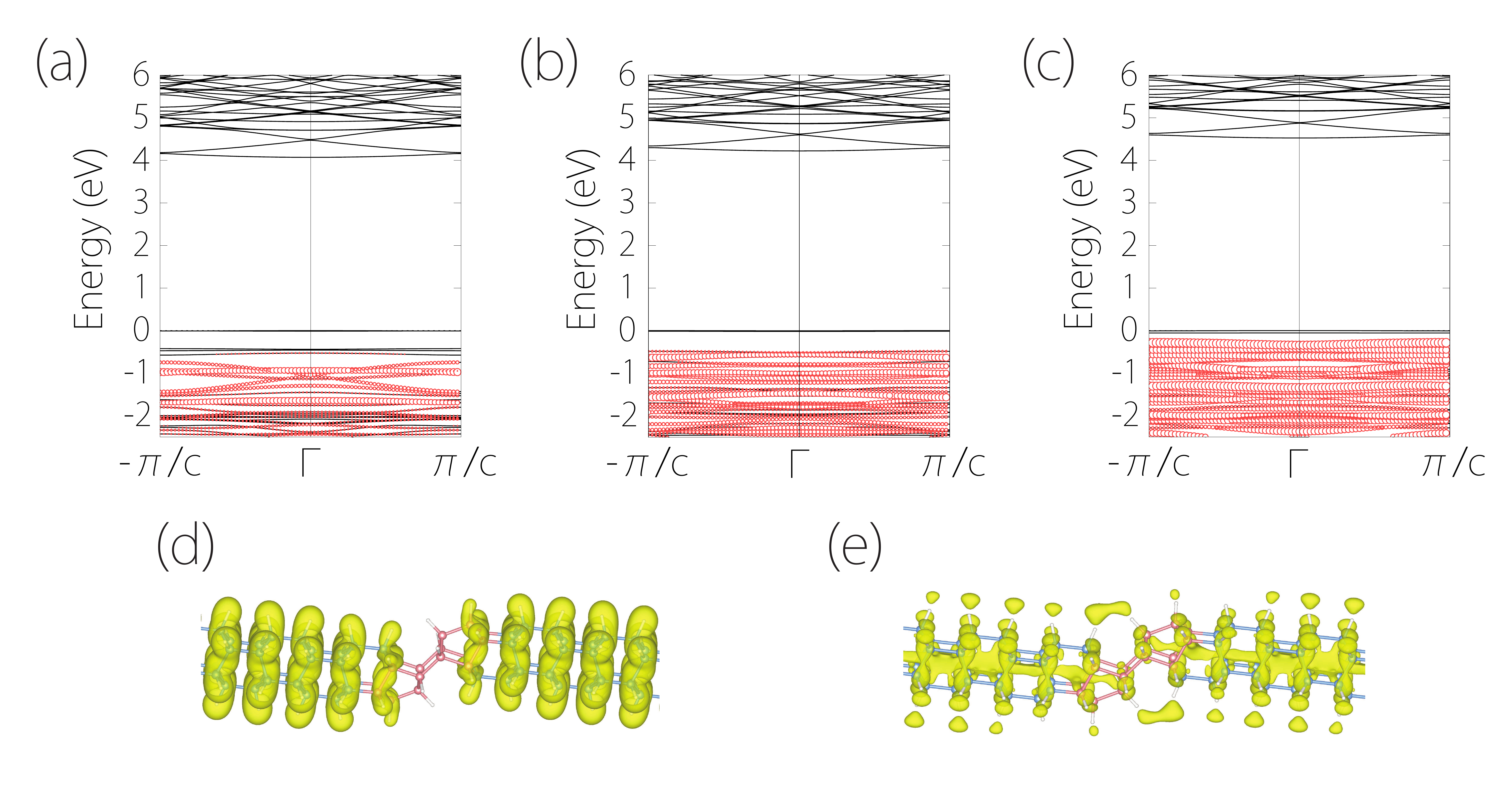}
	\caption{Band structures of SW hybrid nanothreads with (a) single SW defect, (b) double SW defects, and (c) triple SW defects. The Fermi level is set at VBM. The size of the red circles in (a-c) indicates the weight of projection onto the C atoms in the SW defect domains. (d,e) Charge density contour plots for (d) VBM and (e) CBM at the $\Gamma$ point around a single SW defect.}
	\label{fg3:hybrid}
\end{figure}

When SW defects are introduced into the $sp^{3}$-(3,0) nanothread, the bandgap widens as shown in in Table~\ref{tb1:data}. The value increases from $4.07$~eV for the hybrid structure with one SW defect up to $4.82$~eV for the polymer-I nanothread which is fully composed of SW defects. This indicates that SW defects can be used to control the bandgaps of carbon nanothreads. In Fig.~\ref{fg3:hybrid}, we plot the band structures of the SW hybrid nanothreads. As shown in Fig.~\ref{fg3:hybrid}(a), for the structure with one SW defect in a supercell, a quite flat band appears above the original VBM. We analyze the charge density distribution for states of this flat band [see Fig.~\ref{fg3:hybrid}(d)], and find that the flat band mainly involves states distributed in the $sp^{3}$-(3,0) domains, not on the SW defects. Instead, the states distributed on the SW defects contribute to the states below the VBM. This observation implies that the presence of SW defects tend to localize the valence band states in the $sp^{3}$-(3,0) domains. The charge distribution of the CBM state is plotted in Fig.~\ref{fg3:hybrid}(e), from which one observes that the state is mainly distributed inside the thread. The continuous charge distribution is suppressed at the SW defect region. As a result of this enhanced confinement, the presence of defect tends to push up the CBM state in energy, as observed in Fig.~\ref{fg3:hybrid}(a-c). For comparison, in the previous study on $sp^{2}$ (8,0) and (14,0) carbon nanotubes, SW defects are found to enlarge the bandgap size through inducing the localized electronic states in the bandgap, but the gap size decreases with increasing defect concentration~\cite{zhou2014effects}.

\begin{figure}[!htb]
	\includegraphics[width=0.8\textwidth]{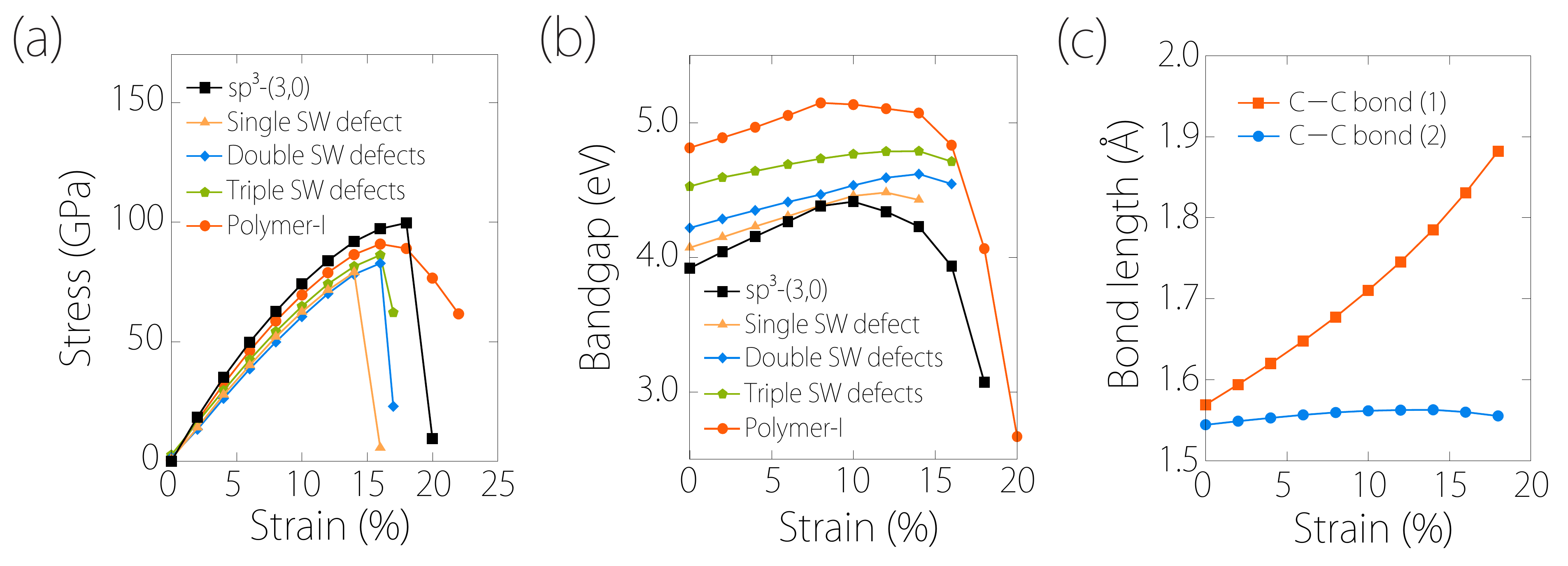}
	\caption{(a) Strain-stress relations and (b) bandgap versus strain curves for the different types of carbon nanothreads. (c) The bond-lengths versus strain for the $sp^{3}$-(3,0) nanothread.}
	\label{fg4:strain}
\end{figure}

Next, we study the strain effects on the electronic properties of carbon nanothreads. Since low-dimensional materials are prone to wrinkle under lateral compression, here we focus on the tensile strain. The strain is defined as $\varepsilon = (l-l_{0})/l_{0} \times 100\%$, where $l$ and $l_{0}$ are the lengths of strained and unstrained structures, respectively. Figure~\ref{fg4:strain}(a) shows the strain-stress relations of the different types of nanothreads. We find that the $sp^{3}$-(3,0) nanothread and hybrid nanothreads exhibit brittle behaviour with the failure strain below ${\sim}20\%$, whereas the polymer-I nanothread can sustain up to ${\sim}22\%$ strain in accordance with the previous work~\cite{Muniz2017DNTmecha}. 

Strain effectively tunes the bandgaps of the nanothreads. From Fig.~\ref{fg4:strain}(b), all nanothreads show a trend of increasing bandgap as the strain increases from zero, but the bandgap drops at larger strain. For the $sp^{3}$-(3,0) nanothread, under strain, the bandgap increases from $3.92$~eV to the maximum of $4.41$~eV at $\varepsilon = 10\%$, while the polymer-I nanothread shows a maximum bandgap of $5.15$~eV at the strain of $8\%$. Notably, for a fixed strain, the bandgap value still increases with the concentration of SW defects. As the number of SW defects increases, the increasing trend of bandgap can be sustained to larger strains, and the slope of bandgap versus strain curve decreases. This shows that the presence of SW defects tends to stabilize the bandgap size against strain.

\begin{figure}[!htb]
	\includegraphics[width=0.8\textwidth]{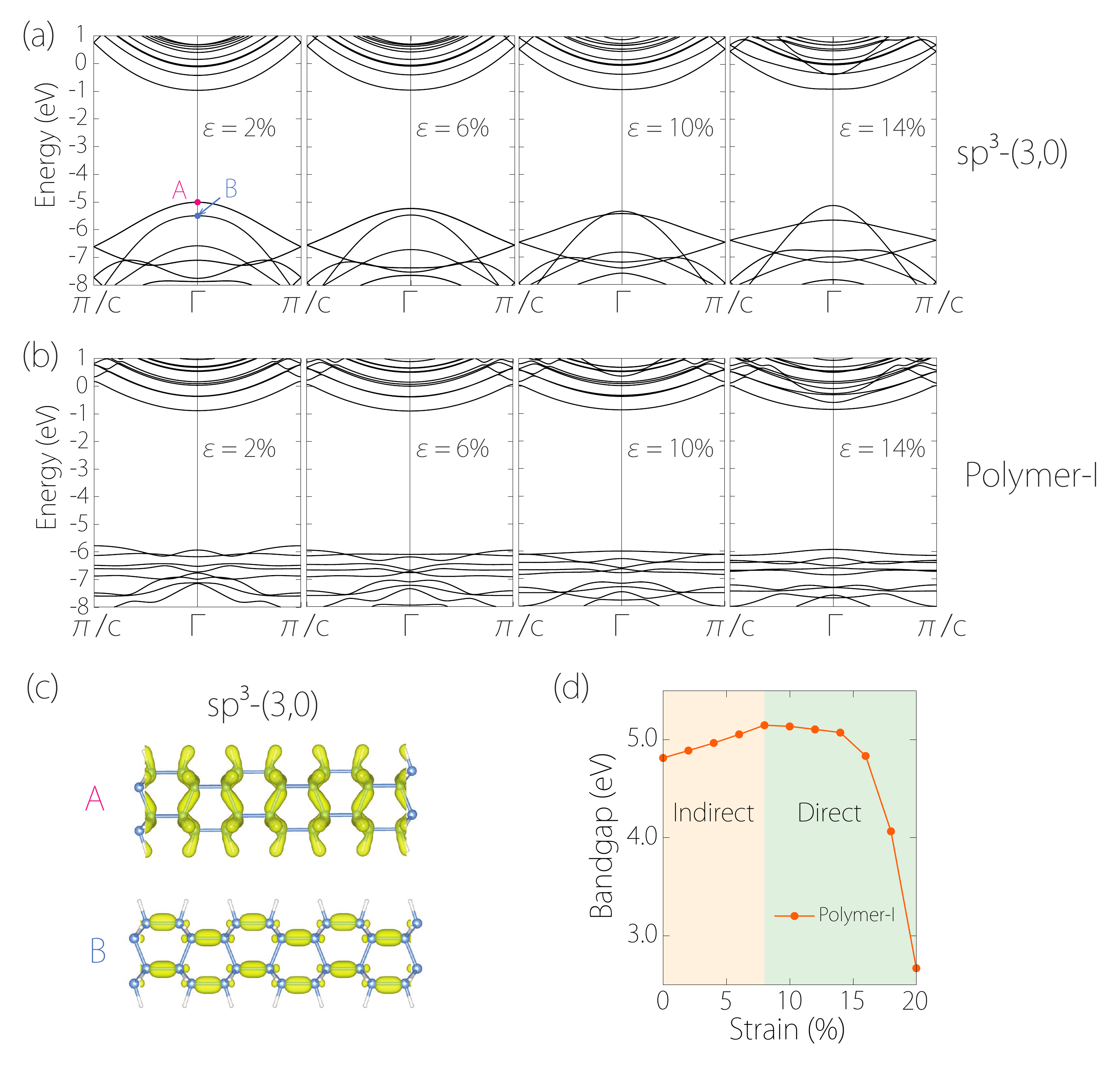}
    \caption{Band structures of (a) $sp^{3}$-(3,0) nanothread and (b) polymer-I nanothread under the strains of $2\%$, $6\%$, $10\%$, and $14\%$. All energies are referenced to the vacuum level. The charge density contour plots of A and B states (as indicated in (a)) at the $\Gamma$ point for the $sp^{3}$-(3,0) nanothread. (d) Indirect-to-direct-bandgap transition in polymer-I nanothread.}
    \label{fg5:bandstrain}
\end{figure}

To further investigate the effects of strain on the electronic properties, we plot the band structures of $sp^{3}$-(3,0) and polymer-I nanothreads under several different strains in Figs.~\ref{fg5:bandstrain}(a) and \ref{fg5:bandstrain}(b). One observes that the CBMs of both $sp^{3}$-(3,0) and polymer-I nanothreads remain almost unchanged. The main change occurs at the VBM. For the $sp^{3}$-(3,0) nanothread, there is an interesting band switching between two bands near VBM at about $8\%$ strain. As indicated in Figs.~\ref{fg5:bandstrain}(a), the original VBM state A is shifted down in energy, causing the increase of bandgap before $\varepsilon = 8\%$. Meanwhile, the B state, initially below A state by ${\sim}0.5$~eV, is shifted up in energy with increasing strain, and at $\varepsilon\approx 8\%$, it crosses the A state and becomes the new VBM.  After the band switching, the B state keeps moving up in energy, leading to the decrease of bandgap. This picture explains the nonmonotonic variation of bandgap value as observed in Fig.~\ref{fg4:strain}(b). 

The different responses of A and B states to strain can be understood from their bonding characters. From the charge distribution of the two states plotted in Fig.~\ref{fg5:bandstrain}(c), one observes that the A state has an antibonding character between the adjacent benzenes, whereas the B state exhibits a bonding character. From Fig.~\ref{fg4:strain}(c), the C-C bond (1) shows continuous increase in length under strain. 
Hence under strain, A state is pulled down whereas the B state is pushed up in energy, which produces the band switching phenomenon. This band switching could lead to dramatic changes in the electronic properties. For example, here the band with A state is flatter than the band with B state. Hence for the case with hole doping, the effective mass of the hole carriers should have a sudden drop when the strain is increased across the critical strain for band switching. 

Similar band switching phenomenon also happens in the polymer-I nanothread between $\varepsilon = 2\%$ and $6\%$ for the two bands at the $\Gamma$ point [see Fig.~\ref{fg5:bandstrain}(b)]. However, the unstrained polymer-I is an indirect bandgap semiconductor with the CBM at $\Gamma$ point and the VBM at $k = \pm \pi/c$. The band switching does not affect the location of VBM. Interestingly, with further increasing strain above $8\%$, the valence band state at the $\Gamma$ point rises above the state at the $k = \pm \pi/c$ point and becomes the new VBM. This causes an indirect-to-direct-bandgap transition induced by strain [see Fig.~\ref{fg5:bandstrain}(d)]. The character of direct/indirect gap is important in determining optical properties of a semiconductor. For example, the intensity of photoluminescence will be much stronger for a direct-gap semiconductor compared with an indirect-gap one. Hence, it is possible to effectively control the optical properties of polymer-I nanothreads by strain.

\section{Conclusion}
Based on first-principles calculations, we have revealed the interesting effects of SW defects and lattice strain on the electronic properties of the newly synthesized carbon nanothreads. The bandgap of nanothreads can be tuned within a wide range of ${\sim}1$~eV by the SW defects in the hybrid structures, interpolating between the bandgaps of $sp^{3}$-(3,0) and polymer-I nanothreads. Under applied strain, the bandgaps of nanothreads show a nonmonotonic variation. The bandgap value first increases with strain, then drops at large strain above 10\%. The gap size can be effectively tuned by strain in a wide range above $>0.5$ eV. Interestingly, for the $sp^{3}$-(3,0) nanothread, a switch of band ordering occurs under strain at the valence band maximum, which can dramatically affect the properties of hole carriers, and for the polymer-I nanothread, an indirect-to-direct-bandgap transition occurs at about 8\% strain, which is expected to strongly affect its optical properties. Since the wide-bandgap diamond has been demonstrated to be a very promising deep-ultraviolet photodetector with solar blindness~\cite{liao2014nanostructured}, given that the carbon nanothreads are of similar bandgap size, they would also offer an encouraging opportunity for applications as photodetectors. Our work suggests that the SW defects and strain offer powerful tools to engineer the properties of carbon nanothreads for future optoelectronic and photoelectrochemical applications as well as stress sensors.

\begin{acknowledgement}
This work is supported by Singapore MOE Academic Research Fund Tier 1 (SUTD-T1-2015004). ZG gratefully acknowledges the financial support from the Agency for Science, Technology and Research (A*STAR), Singapore and the use of computing resources at the A*STAR Computational Resource Centre, Singapore. We also acknowledge the computational support from National Supercomputing Centre Singapore and the Texas Advanced Computing Center.
\end{acknowledgement}

%\begin{suppinfo}
%
%A listing of the contents of each file supplied as Supporting Information
%should be included. For instructions on what should be included in the
%Supporting Information as well as how to prepare this material for
%publications, refer to the journal's Instructions for Authors.
%
%The following files are available free of charge.
%\begin{itemize}
%  \item Filename: brief description
%  \item Filename: brief description
%\end{itemize}
%
%\end{suppinfo}

%\bibliographystyle{achemso}
%\bibliography{manuscript}

\providecommand{\latin}[1]{#1}
\makeatletter
\providecommand{\doi}
{\begingroup\let\do\@makeother\dospecials
	\catcode`\{=1 \catcode`\}=2 \doi@aux}
\providecommand{\doi@aux}[1]{\endgroup\texttt{#1}}
\makeatother
\providecommand*\mcitethebibliography{\thebibliography}
\csname @ifundefined\endcsname{endmcitethebibliography}
{\let\endmcitethebibliography\endthebibliography}{}

\end{document}